# EFFICIENT MARKETS, BEHAVIORAL FINANCE AND A STATISTICAL EVIDENCE OF THE VALIDITY OF TECHNICAL ANALYSIS

Marco Antonio Penteado, Prof. Dr.

(Professor of Technical Analysis, ex-Stock Market Analyst)

**Abstract**

This work tried to detect the existence of a relationship between the graphic signals - or patterns - observed day by day in the Brazilian stock market and the trends which happen after these signals, within a period of 8 years, for a number of securities. The results obtained from this study show evidence of the existence of such a relationship, suggesting the validity of the Technical Analysis as an instrument to predict the trend of security prices in the Brazilian stock market within that period.

(Keywords: Efficient Markets, Behavioral Finance, Technical Analysis)

**Introduction**

For a long time the Efficient Markets Hypothesis - EMH has been an underpinning of the finance theory. However, many critics there appeared stressing that its assumptions did not correspond to the reality of markets. It is still a controversial theme. EMH denies the utility of technical analysis.

Not so long ago Behavioral Finance had its dawning, announcing that a new approach should be given to finance and economics: people, and their behavior, also matter, not only numbers. Also, it came to shed a new light on technical analysis, an instrument that, above all, reflects investor's behavior.

And then, there is Technical Analysis (TA).

This study does not intend to be a course on TA. Rather, we'll introduce concepts necessary to understanding the tests made and the conclusions obtained.

There are several kinds of TA: Bar Charts, Candlesticks, Point & Figure, Elliott Waves. The present study considers only bar charts.

We should stress, however, that TA is an empirical process, i.e., the results come through observation. It is not an exact science, where concepts may be exactly proved or demonstrated.

## 1. THE EFFICIENT MARKETS HYPOTHESIS - EMH

A market is called efficient when prices always fully reflect available information.[1] An efficient capital market is a market that is efficient in processing information. The prices of securities, at any time, are based on "correct" evaluation of all information available at that time.[2] In an efficient market there is a huge number of rational profit-maximizers in active competition trying to predict future market values of individual securities, and the actual price of a security will be a good estimate of its intrinsic value.[3]

Fama[4] defines 3 forms of efficiency: The *weak form*, the *semi-strong* form and the *strong* form: In the *weak form* the information set is historical prices. If the market is efficient in the weak form, no investor is supposed to earn excess (or extraordinary) returns using historical data, once that information is already reflected in prices. Excess returns are profits earned by whatever strategy other than the *buy and hold* strategy and bigger than the profit obtained by this latter strategy. The *semi-strong* form considers information publicly available, like announcements of annual earnings, stock splits, etc. By the same reason, in this form of efficiency an investor couldn't earn excess returns. If the market is efficient in the *strong form* an investor can't earn excess returns with information publicly held or not, even not posted yet.

---

[1] Fama (1970), 383

[2] Fama (1976), 133

[3] Fama (1995), 4

[4] Fama (1970), 383

Fama admits, however, that there is evidence of market inefficiency in the strong form sense, due to the action of market makers, specialists who have a monopoly power over important information and can use it to turn it into profit.[5]

Regarding market efficiency and technical analysis Fama says that *"chart reading, though perhaps an interesting pastime, is of no real value to the stock market investor"*.[6]

The assumption of the EMH that prices reflect all information is criticized by Grossman & Stiglitz.[7] In their opinion, there are costs involved in obtaining information that influence transactions. The authors still argue *"(…) because information is costly, prices cannot perfectly reflect the information which is available, since if it did, those who spent resources to obtain it would receive no compensation. There is a fundamental conflict between the efficiency with which markets spread information and the incentives to acquire information."* In other words, there is an informational asymmetry.

Siglitz still argues[8] *"The most fundamental reason why markets with imperfect information differ from those in which it does (sic) is that actions (including choices) convey information, market participants know this, and this affects their behavior."*

An example of the inefficiency of markets is given by Basu[9], in a study about the predictive power of low P/E ratios. During the period April 1957-March 1971, the low P/E portfolios seem to have, on average, earned higher absolute and risk-adjusted rates of return than the high P/E securities. *"The results suggest a violation in the hypothesis that (…) security price behavior is consistent with the EMH (…) suggesting that P/E ratio information was not "fully reflected" in security prices in as rapid a manner as postulated by the semi-strong form of the EMH."*

---

[5] Fama (1970), 398.

[6] Fama (1965). 34. Fama uses the term *chartist theory* with reference to Dow theory, the granddad of technical analysis. Chart reading or chart analysis may be understood as technical analysis.

[7] Grossman & Stiglitz (1980), 393-408.

[8] Stiglitz (2001), 485.

[9] Basu (1977), 663-682.

Another example of the predictive power of the P/E ratio is given by Shiller[10]. Analyzing the market value of the S&P Composite Index on January, 2000 and dividing it by the 10 years moving average of returns he found a P/E ratio of 44.3, much higher than the 32.6 ratio in September, 1929, on the brink of the crash of the market on Tuesday October 29, 1929. Only as reference, the average P/E ratio of the American stock exchange from 1871 to 1990 had been 15.[11]

Shiller was then heralding a sign of danger[12]: the stock exchange was again on the brink of the burst of a speculative bubble that was being observed in the American stock market.

And he was right, the crisis was near. On April, 14 that year the Nasdaq plummeted, with a drop of 5.66% and the S&P, 5.78%[13]. On April 17 the drop had reached 37% from its highest level on March the same year.

There happened another inefficiency of the market, and the P/E ratio was right. Once again, showing that the semi-strong form of efficiency of the market had failed!

Public data, through the P/E ratio, showed it is possible not only to have excess returns but also to avoid losses.

Information plays a very important role in the concept of the EMH. Bernstein[14] mentions the Victorian concept of rational behavior that *"measurement always dominates intuition: rational people make choices on the basis of information rather than on the basis of whim, emotion or habit. Once they have analyzed all the available information, they make decisions in accord with well-defined preferences."*

Black shows that it isn't always true, arguing that *"people sometimes trade on noise as if it were information.… However, noise trading is essential to the existence of liquid*

---

[10] Shiller (2000), 8.

[11] Newspaper O Estado de São Paulo, 04/15/00 - Brazil

[12] Shiller (2000), 3-14

[13] Newspaper O Estado de São Paulo, 04/15/00 - Brazil

[14] Bernstein (1998), 246.

*markets. People don't know they are trading on noise. They think they are trading on information. Noise is the information that hasn't arrived yet[15]"*

In Electronics, noise is a disturbance superimposed on a useful signal that tends to cloud its informational content.[16] Or, "in its most wide sense, noise is any signal present in a communication channel, other than the desired signal."[17]

Or, as Bernstein puts it, "*noise is badly analyzed information, misinformation or hunch[18]*"

Regarding markets efficiency there are 2 interesting Warren Buffett's quotations:

- *"I'd be a bum on the street with a tin cup if the markets were always efficient"*
  Fortune April 3, 1995

- *"Investing in a market where people believe in efficiency is like playing bridge with someone who has been told it doesn't do any good to look at the cards."*
  http://oldprof.typepad.com June 27, 2006.

For more critics about EMH, see Haugen (1995) and Shleifer (2000).

## 2.  BEHAVIORAL FINANCE

*Behavioral Finance is considered one of the most controversial branches in the New Finance, beside the Fractal Markets Hypothesis (FMH) and the Artificial Neural Networks (ANN).*[19]

Behavioral Finance is a blend of Psychology and Finance, that seeks to explain the (ir)rationality of the decision maker, such as an investor trying to pick one among several stocks to invest.

---

[15] Black (1986). 529; 534.

[16] IEEE (1977), 439.

[17] Bell (1971), 49.

[18] Bernstein (1993)

[19] Ongkrutaraksa (1996).

Thaler[20] arguments that *"Most of the time, the behavior of both rational and less than fully rational agents matter. This means that it is no longer possible to have complete confidence in the usual approach, which is to characterize optimal behavior and then assume this behavior is universal"*. He thinks of behavioral finance as simply *"open-minded finance"*.

### 2.1 Risk Perception

What is risk? Risk may be defined as the degree of uncertainty about an event.[21] Or risk is the probability of the occurrence of failures, where failures are events that keep goals from being achieved, and successes are those that don't. [22]

*"People respond to the hazards they perceive[23]*. But different people perceive risks in different ways[24].

Risk perception leads to judgements, often based on heuristic principles. Heuristics are kind of short-cuts that simplify complex reasoning. The problem is that for the most part heuristics turn reasoning so simple that they lead to errors.

The main heuristics are Representativeness, Availability and Adjustment and Anchoring *"In the representativeness case, probabilities are evaluated by the degree to which A is representative of B, that is, by the degree to which A resembles B. (…) This approach to the judgement of probability leads to serious errors because similarity, or representativeness, is not influenced by several factors that should affect judgements of probability.[25]* In many situations people assess the probability of an event or the frequency of a class by the ease with which instances or occurrences can be brought to mind. This is the case of the heuristic called availability. Oftentimes people make estimates by starting from an initial value that is adjusted to yield the final answer. This reference, or starting point is, many times, wrong, leading to insufficient adjustments. This is the case of adjustment and anchoring[26] which is, in our opinion, the most harmful heuristic; for instance, when the price of a security starts to fall and after some time, with prices still falling, an investor decides to buy in order to improve the average price. That proceeding usually leads to losses.

---

[20] Thaler (1993). xvi; xvii

[21] Solomon & Pringle (1981, *apud* Securato 1996), 21.

[22] Securato (1996), 28

[23] Slovic *et al*, (1980, *in* Kahneman *et al* 2001), 463

[24] Weber & Hsee (1988), 1205-1217.

[25] Kahneman & Tversky (1974, *in* Kahneman *et al* 2001), 4

### 2.2 The Prospect Theory[27]

In our opinion that is the most important point of Behavioral Finance.

Imagine 2 mountains, one much bigger than the other. The bigger one should seem always bigger, it should be an invariance. But that is not what happens. Suppose you are driving down a road and you are close to the smaller mountain, and the bigger mountain is very far away. The impression you have is that the mountain that is near is bigger than the one that is very far. That is a failure of invariance.[28]

That kind of distortion often happens when you change the framing of a question: different ways of proposing a question lead to different answers, or, different frames of presenting a problem may lead to different decision making.

In a nutshell, people are risk averse in choices involving sure gains and risk seeking in choices involving sure losses.

People react to the risk they perceive.

People don't perceive risk in the same way

People don't perceive risk at the same time.

Losses loom bigger than gains.

---

[26] *Idem*, 14.

[27] Kahneman & Tversky (1979)

[28] Bernstein (1998), 269-283.

# 3. TECHNICAL ANALYSIS

## 3.1. Introduction

What is Technical Analysis? TA may be considered a mix or a combination of art and science, according to the vision of many authors. "It is one approach to marketing forecasting based on a study of the past, human psychology…)"[29]. It is the science of recording, usually in graphic form the (…) price history of a certain stock then deducing the probable future trend.[30] History repeats itself, and prices move in trends. Technical Analysis is based on following trends.[31] It is the study of repeating patterns which reflect investors' waves of greed and fear.[32]

## 3.2. Dow Theory

TA, as it is known today, stems from Dow Theory, which can be considered TA's granddad. Charles Dow and Edward Jones founded the Dow Jones & Co. in 1882, in order to post stock prices and economic news from the New York stock market. Dow was the founder of the Wall Street Journal, and they were the journal first editors[33]. Dow discovered that stock prices do not move in a chaotic way, but rather, follow trends established by the stock market as a whole.[34] The first publication of a stock market average happened on July 3, 1884, and that index included only 11 stocks, nine of which were railroad companies.[35]

Charles Dow never wrote a book himself, but his ideas, known through his editorials, were organized by William Peter Hamilton in his book *The Stock Market Barometer* and first mentioned as Dow Theory by Samuel Armstrong Nelson in his book *The ABC*

---

[29] Murphy (1986), xviii.

[30] Edwards & Magee (2001), 4.

[31] Goldberg & Nitsch (2001), 6

[32] Noronha (1995), 1.

[33] Hamilton (1922), 4.

[34] Leite & Sanvicente (1995), 12.

[35] Murphy (1986), 24.

*of Stock Speculation*. Hamilton's editorials are reproduced in Rhea's book *The Dow Theory*. These 3 books were edited after Dow's death.

Among Dow's tenets there is one that, in our opinion, is the most important of all and says: ***a trend is still valid until a signal of its reversal is confirmed.*** In other words, ***after a trend has started it is more likely to continue or keep its direction than to reverse.***[36]

### 3.3 The Bar Chart

The bar[37] in the bar chart (fig. 1) represents a trading period, i.e., a day, a week, one hour, one minute, 5 minutes and so on

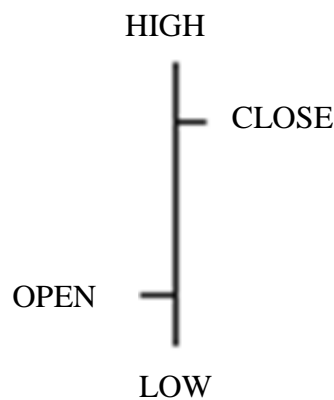

Fig.1  THE BAR

OPEN – the price of the first trade in the period,

CLOSE – the price of the last trade in the period

HIGH – the highest price in the period

LOW – the lowest price in the period

---

[36] Murphy (1986), 3; 30. Rhea (2002), 76.

[37] Achelis (2001), 9.

**The number Three**

One of the features of TA is the importance of the number 3, as we shall see along the study. It's kind of a pattern, to be observed in many situations.

**Tops and Bottoms**

In the stock market there is a saying that goes "buy cheap and sell dear". But how much is cheap and how much is dear? To buy cheap means you should buy at the bottom; and to sell dear, that you should sell at the top. If it weren't by pure luck, you'll only know that a point is an extreme, top or bottom, after you have passed it. You'll only know that you reached the top of the mountain when you start to descend the other side. In terms of TA it means that to confirm that one bar is the top it is necessary to have at least 3 bars going down after it; or that it is the bottom, at least 3 bars going up after it. See fig.2 and 3.

**TOP or High**

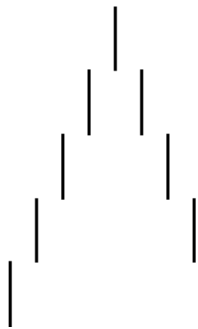

Fig. 2 – Confirmation of a top or a high

**BOTTOM or Low**

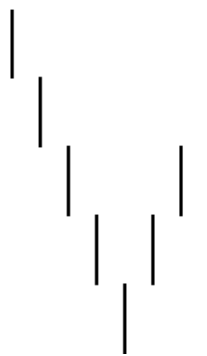

Fig. 3 – Confirmation of a bottom or a low

**Trendlines**

Trendlines indicate the main direction of prices evolution, and can be Up Trendlines or Down Trendlines (see fig. 4 and 5).

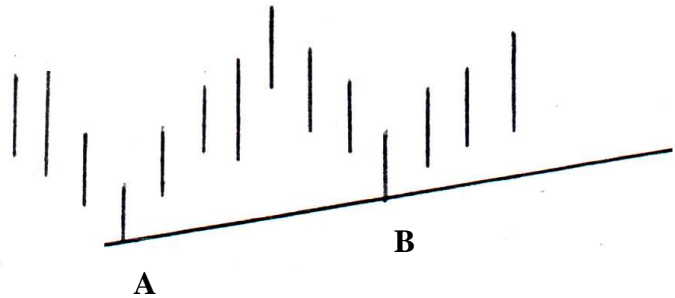

Fig. 4 – Up Trendline

To trace an up trendline you need 2 lows (bars A and B), low B > low A, separated by at least 3 bars.

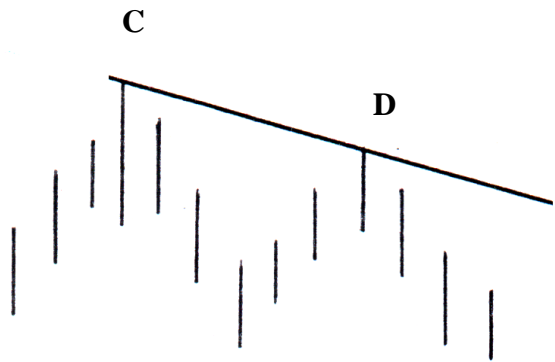

Fig. 5 – Down Trendline

To trace a down trendline you need 2 highs (bars C and D), high C > high D, separated by at least 3 bars.

Thus the definition of a trend uses to take at least 8 days.

**Supports and Resistances**

Support is a limit below which prices (or points) don't fall in movements above an up trendline (fig. 6A) or above a horizontal line in sideways movements (fig. 6B).

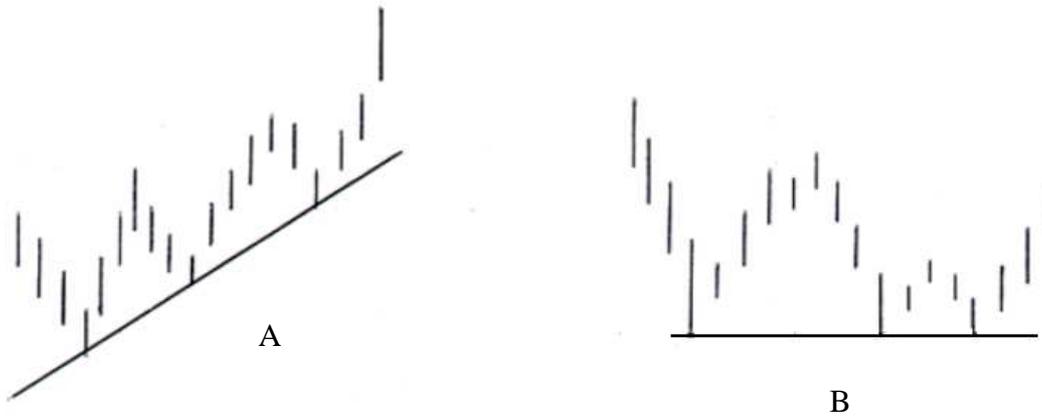

Fig. 6 - Supports

Resistance is a limit that is not passed in up movements below a downtrend (fig. 7A) or below a horizontal line in sideways movements (fig. 7B).

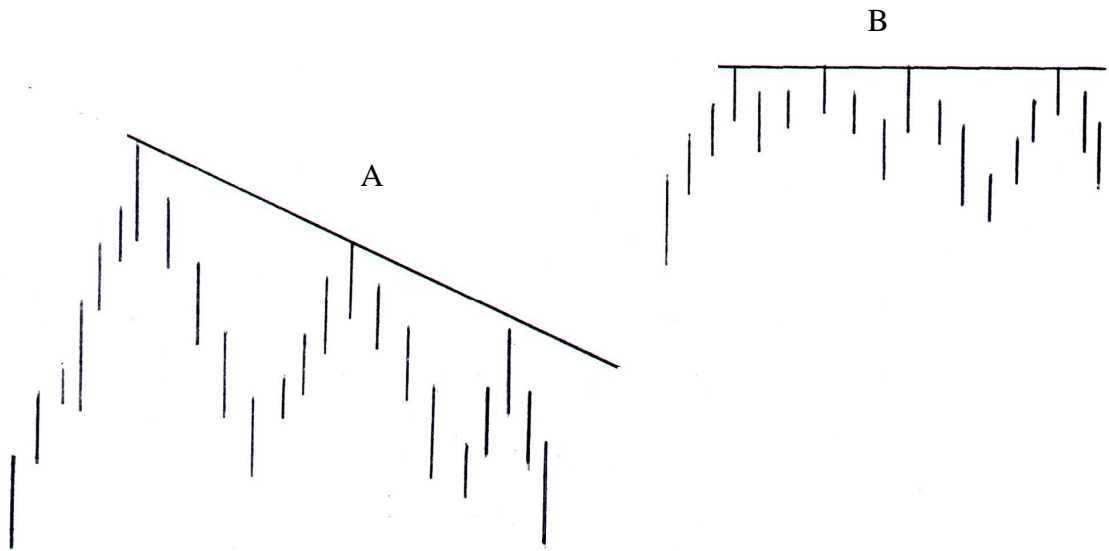

Fig. 7 – Resistances

## Trend Reversal – Breaking Resistance/Support levels[38]

To revert, a trend must break the resistance/support level by at least 3%, i.e., the closing of the bar on that day should be at least 3% above/below that level. If it doesn't reach 3% on a single day you just add the ups/downs in the two following days. The breakthrough is also valid if there are 3 bars going up/down after the crossing point even if the percentage is less than 3%, as shown below. Three days or 3% apply indistinctly to horizontal lines or up/down trendlines.

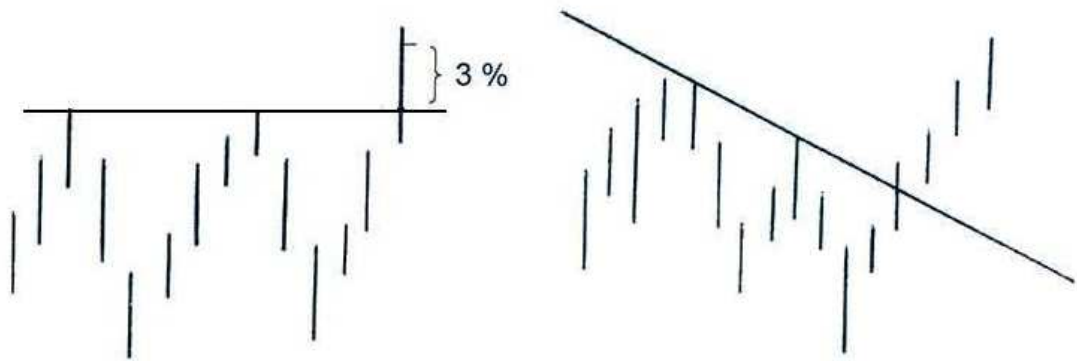

Fig. 8 - Breakthrough

## Moving Averages

A moving average (MA) is kind of a moving window where you can see a fixed number of prices. For instance, in an arithmetic moving average (AMA) of 5 days you add the prices of the last 5 days and divide the result by 5. Each new day you include this most recent day and discard the first, 5 days ago.

In a brief explanation, its value is given by the formula

$$AMA = \frac{\sum_{1}^{n} p_i}{n}$$

Where $p_i$ is the price $p$ on day $i$, and $n$ is the number of days considered (window size).

---

[38] See: Noronha (1995), 54. Murphy (1986), 66-67; 74. Edwards & Magee (2001), 57. Pring (2002), 73.

Moving averages behave like resistance/support areas. When prices cross a MA it's a signal that the trend is starting to revert. When prices cross above a MA and prices and MA spread out, confirming, it's a buy signal. When they cross below and prices and the MA spread out confirming, it's a sell signal.

The most usual MAs are MA 21 days (representing 1 month, usually 21 working days); MA 233 days (representing one year, usually 252 working days).[39]

MA 233 usually is the most important resistance/support level.

**4. THE RESEARCH**

In order to show the validity of technical analysis in the Brazilian Stock Market through a statistical analysis of the occurrence of graphic signals (patterns) and its relation with the subsequent trends, this study considered the day to day evolution of prices of 10 securities – the BOVESPA index and 9 stocks from the index: Petrobras PN, Eletrobras PN, Vale PNA, Bradesco PN, Cemig PN, Itau PN, Usiminas PNA, Siderurgica Nacional ON e Tubarao PN, within the 8 years period from 1995/01/02 to 2002/12/30.

The patterns considered were:.

Up movements: up trendlines, breaking of down trendlines, breaking of horizontal resistances, double bottoms, flags, channel, high fan principles, symmetrical triangle, rectangle, head and shoulders, bullish pennants.

Down movements: down trendlines, breaking of up trendlines, breaking of horizontal supports, down fan principles, symmetrical triangle, head and shoulders, flags, bearish pennants, double tops

**5. METHODOLOGY**

The goal of this study is to check the validity of the patterns used by the technical analysis. For instance: I buy, for a certain price, a stock in the next day an up trendline has been confirmed[40] and sell it the next day its breaking is confirmed[41]. Once the selling price is bigger than the buying price, I'll have a profit. Therefore, the trendline is true. Otherwise, it's false. After selling, the process starts all over again. Let us make it clear that the trend – that is established after its confirmation 3 days after the second low or high, constitutes the pattern itself.

---

[39] Those numbers of days come from the sequence of Fibonacci; see Brown (2008) and most of the references already mentioned.

[40] See p.11

This process was used on a daily basis, starting in the first day of the period – 95/01/02 until its last day, 02/12/30 and repeated for each one of the securities above mentioned, all patterns were registered and analyzed (see Table 1, below). The total was tested through Chi Square test, showing statistical relevance at 0.5% level of significance.

## 6. THE TESTS

| PATTERN | TOTAL | % | % accum. | TRUE | FALSE |
|---|---|---|---|---|---|
| Up Trendline | 203 | 24.2 | 24.2 | 153 | 50 |
| Breaking of a Downtrendline | 183 | 21.8 | 46.0 | 133 | 50 |
| Breaking of an Up Trendline | 182 | 21.7 | 67.6 | 131 | 51 |
| Down Trendline | 153 | 18.2 | 85.8 | 117 | 36 |
| Breaking of a Horizontal Resistance | 26 | 3.1 | 88.9 | 15 | 11 |
| Double Bottom | 19 | 2.3 | 91.2 | 19 | 0 |
| Breaking of a Horizontal Support | 14 | 1.7 | 92.9 | 10 | 4 |
| Double Top | 13 | 1.5 | 94.4 | 11 | 2 |
| Up Flag | 7 | 0.8 | 95.2 | 6 | 1 |
| Down Fan Principle | 7 | 0.8 | 96.1 | 7 | 0 |
| Channel | 6 | 0.7 | 96.8 | 6 | 0 |
| Down Head & Shoulder | 6 | 0.7 | 97.5 | 6 | 0 |
| High Fan Principle | 5 | 0.6 | 98.1 | 3 | 2 |
| Up Symmetrical Triangle | 5 | 0.6 | 98.7 | 5 | 0 |
| Retangle | 4 | 0.5 | 99.2 | 4 | 0 |
| Up Head & Shoulders | 2 | 0.2 | 99.4 | 2 | 0 |
| Down Flag | 2 | 0.2 | 99.6 | 2 | 0 |
| Up Pennant | 1 | 0.1 | 99.8 | 0 | 1 |
| Down Pennant | 1 | 0.1 | 99.9 | 1 | 0 |
| Down Symmetrical Triangle | 1 | 0.1 | 100.0 | 1 | 0 |
| TOTAL | 840 | 100.0 | | 632 | 208 |
| | | 100.0 % | | 75.2 % | 24.8 % |

Table 1 – Results (Ups and Downs, Consolidated)

The results in table 1 show that 75.2% of the total patterns obtained were true, compared with only 24.8% false, thus indicating the reliability of technical analysis.

---

[41] See p.13

Table 1 also shows that the first four patterns – up trendline, breaking of a down trendline, breaking of an up trendline and down trendline, the most frequent patterns, account for nearly 86% of the total of patterns and are the most significant patterns. Results also tested through Chi Square test, showing statistical relevance at 0.5% level of significance.

| PATTERN | TRUE OR FALSE | | | | |
|---|---|---|---|---|---|
| | TOTAL | TRUE | % | FALSE | % |
| Up Trendline | 203 | 153 | 75.4 | 50 | 24.6 |
| Breaking of a Downtrend | 183 | 133 | 72.7 | 50 | 27.3 |
| Breaking of an Up Trendline | 182 | 131 | 72.0 | 51 | 28.0 |
| Down Trendline | 153 | 117 | 76.5 | 36 | 23.5 |

Table 2 – Relevance of the most important patterns

Table 3 shows a very interesting point, that each security has its "own personality". Bradesco PN presented the highest percentage of true patterns, 88.3%. It means that whenever an up pattern was confirmed in Bradesco PN chart an investor should buy it with a chance of 88.3%, on average, of making money. In spite of also presenting a high percentage, Usiminas PNA is the worst among all of them, with 68.3% of true patterns.

*"All stocks vary with the market as a whole, but each responds to information in its own way"*[42]

| SECURITY | TRUE PATTERNS | % | FALSE PATTERNS | % | TOTAL |
|---|---|---|---|---|---|
| BRADESCO PN | 83 | 88.3 | 11 | 11.7 | 94 |
| VALE PNA | 59 | 81.7 | 13 | 18.3 | 72 |
| SID. TUBARÃO PN | 54 | 79.4 | 14 | 20.6 | 68 |
| SID. NACIONAL ON | 59 | 77.6 | 17 | 22.4 | 76 |
| CEMIG PN | 57 | 74.0 | 20 | 26.0 | 77 |
| ÍNDICE BOVESPA | 76 | 73.8 | 27 | 26.2 | 103 |
| ITAÚ PN | 79 | 72.5 | 30 | 27.5 | 109 |
| ELETROBRAS PNB | 46 | 68.7 | 21 | 31.3 | 67 |
| PETROBRAS PN | 63 | 68.5 | 29 | 31.5 | 92 |
| USIMINAS PNA | 56 | 68.3 | 26 | 31.7 | 82 |
| TOTAL | 632 | 75.2 | 208 | 24.8 | 840 |

Table 3 – Results per security

---

[42] Bernstein (1993), 258.

## 7. CONCLUSION

The goal of this research was to show the validity of technical analysis in the Brazilian stock market through the study of graphic patterns detected within the 8 years observation period, in the selected securities.

Results show that from the 840 identified patterns, 632 were true (75.2%) and 208 (24.8%), false, as to validate the following trends, i.e., this study has shown that true patterns constitute the most of the patterns encountered, thus validating the efficiency of technical analysis, and so suggesting inefficiencies of the market and contradicting the EMH in its weak form when it affirms that historical data are already reflected in prices, also showing that the affirmation *"chart reading, though perhaps an interesting pastime, is of no real value to the stock market investor"* may not correspond to reality.

Stiglitz won the Economy Nobel prize in 2001 with the theory of Informational Asymmetry (see p. 3). Technical Analysis is costly information, you have to pay for it, what causes an asymmetry.

Kahneman won the Economy Nobel prize in 2002 with the theory of Behavioral Finance. Technical Analysis reflects behavior.

These two Academy awards reinforce the importance of Technical Analysis.

Basu (see p. 3) indicated inefficiencies in the EMH in its semi-strong form.

Fama himself admitted inefficiencies in the EMH in its strong form.

This study thus indicates the strengthening of behavioral finance, the questioning of efficient markets hypothesis and the validity of technical analysis.

*"I'd be a bum on the street with a tin cup if the markets were always efficient".*

*Warren Buffet,* Fortune April 3, 1995